\documentclass[useAMS,usenatbib]{mn2e}
\usepackage{epsfig,amssymb,amsmath,rotating}

\title[Discovery of one new class in GRS 1915+105]{Discovery of one new class in the light curve of GRS 1915+105}
\author[Pahari et al.]{Mayukh Pahari$^{1, 2, 3}$\thanks{E-mail: mp@tifr.res.in (MP), spal@ncra.tifr.res.in (SP)} and Sabyasachi Pal$^{2}$\\
$^{1}$Department of Physics, University of Pune, Pune, 411007, India\\
$^{2}$National Centre for Radio Astrophysics, Pune, 411007, India\\
$^{3}$Tata Institute of Fundamental Research, Mumbai, India}
\begin{document}

\date{}

\pagerange{\pageref{firstpage}--\pageref{lastpage}} \pubyear{2002}

\maketitle

\label{firstpage}

\begin{abstract}
From the study of X-ray light curve and color-color diagram of the low mass X-ray binary GRS 1915+105, observed by on board proportional counter array (PCA) of Rossi X-ray Timing Explorer (RXTE), we discover a new class of variability, which we name $\epsilon$ class. We have studied observations between MJD 51200 and 51450. The class shows unusual periodic-like variation in count rate during rise time of two x-ray bursts. The class take place when the source is in radio quiet state. 
The huge expansion in color-intensity diagram indicates the class to be an adjusting stage of increasing accretion rate. Spectral analysis shows that during lower count rate, the spectrum is hard power-law dominating, indicating similarity towards hard intermediate state, and during higher count rate, the spectrum is thermal disk blackbody component dominating, indicating similarity towards high soft state. Hence, this class is important in understanding the way of state transition leads to change in accretion rate.
No signature of any low frequency quasi periodic oscillation was seen in this class. We also find that when the class was showing higher counts, the average RMS amplitude is significantly high for high energy band (14-60 keV) compared to low energy band (2-8 keV).
\end{abstract}

\section{Introduction}
GRS 1915+105 is one of the well observed low mass X-ray binary and was discovered by WATCH all-sky monitor on board Granat satellite in 1992 \citep{Ca92}. The source is one of the brightest X-ray source in the sky and shows superluminal ejections \citep{Mi94} when the velocity of the radio jet is more than $0.9c$. GRS 1915+105 is a black hole with $14.0\pm4.4 M_\odot$ \citep{Ha04} which is consisting of a K-M III binary star of $0.8M_\odot$ in a wide 33.5 day orbit \citep{Gr01}. The source is situated $\sim12$ kpc far \citep{Mi94,Fe99}. The GRS 1915+105 shows considerable variability in its light curve and \citet{Be00} showed that different patterns in light curve and hardness ratio of the source is repeating and classified the light curve of the source in 12 separate classes, namely $\alpha$, $\beta$, $\gamma$, $\delta$, $\theta$, $\kappa$, $\lambda$, $\mu$, $\nu$ $\rho$, $\phi$ and $\chi$. Later discovery of another two classes of variability were reported \citep{Na02,Ha05} which are not so common like other 12 classes. \citet{Be00} also phenomenologically classified X-ray states of GRS 1915+105 into three basic states, two softer states namely A, B, when the full accretion disk is observable and one harder state C when the innermost part of the accretion disk is not observable. The source spends most of the time in relatively hard state C. Here we are reporting discovery of one new class from the study of light curve and color-color diagram of the source, observed by Rossi X-ray Timing Explorer (RXTE). We are calling the new class as $\epsilon$ class.

\section{Observations and data analysis}
RXTE is regularly observing GRS 1915+105 from 1995 \citep{Gr96,Be00}. We have analyzed RXTE proportional counter array (PCA) data starting from MJD 51200 to MJD 51450 which includes total 97 observations. The data were reduced using LHEASOFT version 6.6.3. We used the standard convention of good time interval (exclusion time for the South Atlantic Anomaly (SAA) of 30 minutes, offset pointing $< 0.02^o$ and elevation above earths limb$> 10^o$) for our observation. 
We first extracted light curves in three different energy bands to produce hardness ratio, color-color diagrams and hardness
intensity diagrams using cleaned 1-s resolution data. We define 3 bands as 2.0-5.8 keV (X), 6.0-13.6 keV (Y) and 14.0-60.0 keV (Z) and colors are defined as $HR1=Y/X$ and $HR2=Z/X$. For extracting light curve of different bands, we used same number of PCUs for all the observations. To probe in details the characteristics of the source during high and low count rate, we separated the $\epsilon$ class light curve in two distinct types where {\it type I} is characterized by higher count rate and {\it type II} has lower count rate.
We also study the hardness-intensity diagram by studying count rate in three different energy bands.  In  hardness-intensity diagram we have plotted HR2 vs Total flux $(X+Y+Z)$ in three bands. To investigate any presence of low-frequency quasi-periodic frequency (LQPO) between 0.1-10 Hz, we produce power-density spectrum by using {\it powspec}. The power spectrum is normalised such that their integral gives the squared rms fractional variability and the expected white noise level is also subtracted. We use 10 milli sec. time-resolved data with geometrical binning of -1.06 for the study of power-density spectrum. 
To study variation of rms amplitude, we extracted light curves between 2.0-8.0 keV and 16.0-60.0 keV using binned data having 0.1 sec time resolution. From these, we have calculated RMS amplitude for these two different energy bands for each 1 sec binning.

\begin{figure}
\includegraphics[scale=.40]{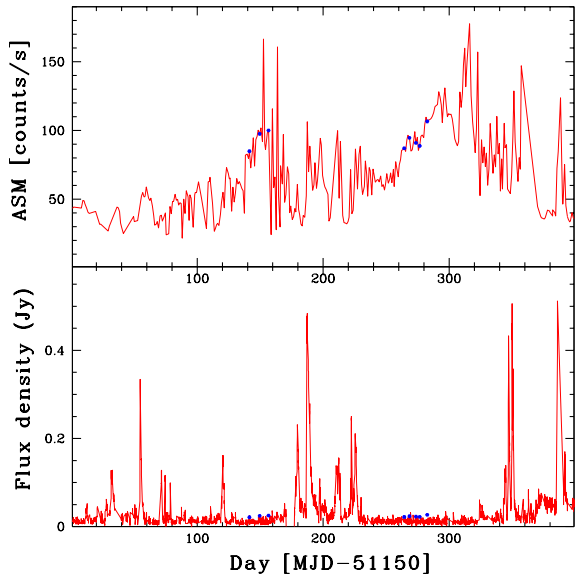}
\label{asm_gbi}
\caption{({\it Upper panel:}) RXTE All Sky Monitor Light curve, at the time of the occurrence of this class. ({\it Lower panel:}) Green Bank Interferometer (GBI) radio flux density of the source in the same time. In both of these figures the time of new class is marked in blue color. }
\end{figure}

We have extracted spectrum from PCA as well as HEXTE data. For PCA data reduction, we have used all the cleaned data file having 16s resolution and corresponding response matrix file has been created adding all xenon layers of PCU 0, 1 and 2 using {\it pcarsp}. Then from the same data file we have extracted background spectrum using {\it runpcabackest}. For HEXTE data reduction, we have extracted source and background file using {\it hxtback}. Dead time correction has been applied to those spectrum files, then response matrix file is created from source spectrum using {\it hxtrsp}. Three PCA and three HEXTE files are fed to XSPEC for fitting. After background subtraction, we found that the HEXTE background subtracted data is only about 13 percent of un-subtracted spectrum and highest HEXTE count rate is less than 1\% of highest PCA count rate found in spectrum. Since the count rate in HEXTE data is very low, errors are very high and so for spectral fitting, we did not use HEXTE data. Since the PCA count rate is low towards very low ($<2$keV) and very high energy end ($>$30 keV), we have also ignored those energy for spectrum fitting and effective energy range of $3-30$ keV was used for fitting. 1\% systematic error has been included. We have separately fitted data for {\it type I} and {\it type II}. We have chosen standard model of black hole consisting of powerlaw, multi-colored disk blackbody component \citep{Ku98} and a small gaussian to take care of 6.4 keV Fe line in accretion disk. For {\it type II}, a single power law and {\it diskbb} was needed to fit the spectrum but for {\it type I,} a broken power law was needed along with disskbb. These additive components which are sole properties of the disk has been modified by interstellar absorption with column density. For two-component model fitting, the value of $N_H$ is found to be 6 to 8 $\times 10^{-22}$ with 95\% confidence level. Hence, we have fixed the column density to $7 \times 10^{22}$ cm$^{-2}$ for uniformity. 

\section{The new class $\epsilon$}
In the upper panel of Figure 1
, RXTE All Sky Monitor (ASM) light curve, at the time of the occurrence of this new class is plotted and in the lower panel of same Figure Green Bank Interferometer (GBI) radio flux density of the source is plotted for the same time. In both of these figures the time of new class is marked in blue color. It is found that the new class took place when the source was in radio quiet state, weeks before a radio flare. The source was in a X-ray flare for all the time when the class was observed and always it was observed in the rising part of the flare.

\subsection{Timing features of the new class}

\begin{figure}
\includegraphics[scale=.40]{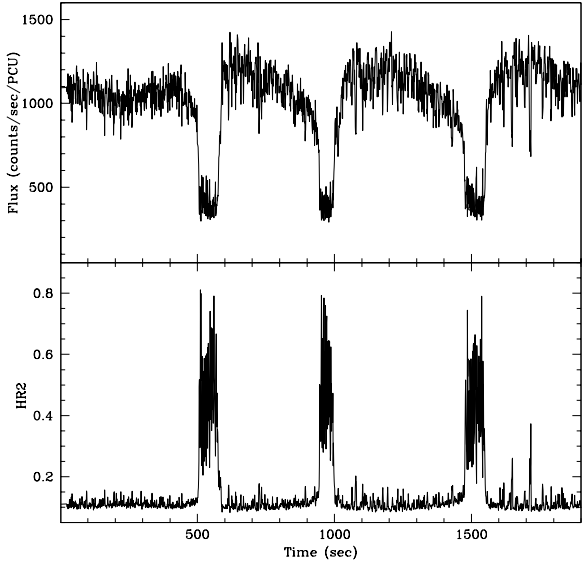}
\caption{{\it Upper panel:} A typical lightcurve of GRS 1915+105 in newly discovered $\epsilon$ class in the energy range 2-60 keV. The observation took place on 23rd April 1999.{\it Lower panel:} Variation of HR2 for the same time as in upper panel.}
\label{light}
\end{figure}

\begin{figure}
\includegraphics[scale=.40]{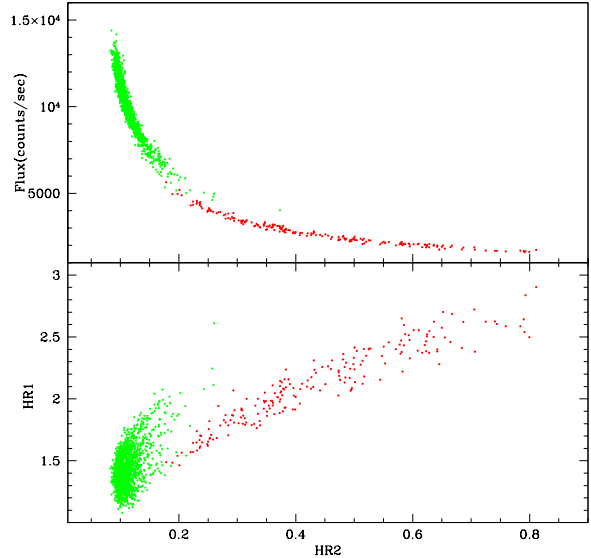}
\caption{{\it Upper panel:} {\bf hardness-intensity diagram}, showing variation of total flux with HR2 for the data observed on 23rd April 1999.{\it Lower panel:} {\bf color-color diagram}, showing variation of HR1 with HR2. In both of these figures, points corresponding to {\it type I} is plotted in green and points corresponding to {\it type II} is plotted in red. Points corresponding to {\it type I} and {\it type II} form two clearly distinguishable separate region on both of the plots.}
\label{color}
\end{figure}

The new class is characterized by nearly periodic fluctuation between higher and lower count rate where the higher count rate is about 3-4 times more than the lower count rate. In the upper panel of the Figure \ref{light}, a typical light curve of GRS 1915+105 at the time of $\epsilon$ class is shown, which was observed on 23rd April 1999. The source shows higher count rate for 350 to 500 s followed by lower count rate of 80 to 100 s. This indicates some kind of thermal-viscous instability in the disk which is adjusting itself to accommodate higher accretion rate at the time of X-ray flare. In the lower panel of the Figure \ref{light}, variation of HR2 with time is shown. It is found that at the time of {\it type II}, HR2 increases to $\sim 0.5$ from $\sim 0.1$ of {\it type I} and thus while lightcurve is varying between {\it type I} and {\it type II}, the source is undergoing an oscillatory state transition between soft and hard states. In the upper-panel of Figure \ref{color}, we plotted color-intensity diagram, where the variation of total flux with HR2 is shown. In the lower panel of Figure \ref{color}, we have shown color-color diagram, where variation of HR1 with HR2 is shown. 
In color-color diagram and hardness-intensity diagram, points corresponding to {\it type I} and {\it type II} are shown in green and red color respectively. In these diagram {\it type I} and {\it type II} form two clearly separable regions. Since the intensity in {\it type I} is high and HR2 value is low, it is in high-soft state. The {\it type II} region is in the hard-intermediate state.
In Figure 4
, the power-density spectrum of the source is shown. No, definite low-frequency quasi-periodic oscillation (LQPO) has been observed. After fitting with lorenzian between the frequency range of 0.01 to 10 Hz, we calculated total power 0.016 for {\it type I} and 0.116 for {\it type II}. Total power of {\it type I} state is $\sim 7$ times lower than the total power of {\it type II}, this result is also expected if we assume {\it type I} to be soft state and {\it type II} to be hard state.
From our analysis it is clear that the {\it type II} state is mostly similar to state `C' while the {\it type I} state is mostly similar to state `B' where definition of the staes are same as mentioned by \citet{Be00}. So, it is clear that transition from state `C' to state `B' and and state `B' to state `C' both are possible. This class is special as in the previous study with all earlier 12 classes \cite{Be00} demanded that the transition from state `C' to state `B' is not possible and $\epsilon$ is the only class, so far discovered, where this kind of transition is possible. 

\begin{table*}
\centering
\caption{Observation ID and Spectrum fitting parameter for {\it Type I}}
\begin{tabular}{cccclcccccc}
\hline
Obs. ID&Date&$\Gamma_1$&$\Gamma_2$&Break&$T_b$&$\chi^2/f$&$F_{2-26}$&$F_{pow}$&$F_{blackbody}$&$F_{iron}$\\
&&&&Energy (keV)&(keV)&&&&&\\
\hline
40403-01-07-00& 04/23/1999 &3.42$\pm$0.06 &4.81$\pm$0.06 &18.3$\pm$0.3 &1.95$\pm$0.01 &1.03 &2.16 &0.81 &1.35 &0.004\\
40703-01-13-01& 05/01/1999 &3.58$\pm$0.07 &4.87$\pm$0.06 &17.5$\pm$0.3 &1.99$\pm$0.02 &0.96& 3.22& 1.64& 1.57& 0.006\\
40703-01-14-00& 05/08/1999 &3.21$\pm$0.04 &4.77$\pm$0.06 &16.7$\pm$0.3 &1.90$\pm$0.00 &0.92 &2.77 &1.25 &1.52 &0.004\\
40703-01-27-01& 08/23/1999 &3.11$\pm$0.06 &4.91$\pm$0.10 &16.0$\pm$0.3 &1.94$\pm$0.01 &0.96& 1.82& 0.74& 1.09& 0.001\\
40703-01-27-00& 08/23/1999 &3.54$\pm$0.05 &4.92$\pm$0.11 &18.1$\pm$0.4 &1.97$\pm$0.01 &1.03 &2.41 &1.07 &1.33 &0.006\\
40703-01-28-02& 08/28/1999 &3.33$\pm$0.06 &4.88$\pm$0.11 &19.0$\pm$0.4 &2.01$\pm$0.02 &0.85& 1.71& 0.59& 1.12& 0.004\\
40703-01-29-00& 09/05/1999 &3.28$\pm$0.04 &4.77$\pm$0.06 &17.8$\pm$0.3 &1.92$\pm$0.01 &1.06 &2.66 &1.25 &1.40 &0.005\\
\hline
\end{tabular}
\flushleft Column density $\sim7 \times 10^{22}$ cm$^{-2}$ was used. Flux is given in the unit of photon (1 photon$ = 9.012 \times 10^{-9}$ ergs/cm$^{-2}$/sec in 2-26 keV band).
\end{table*}
  
\begin{figure}
\includegraphics[angle=-90,scale=.350]{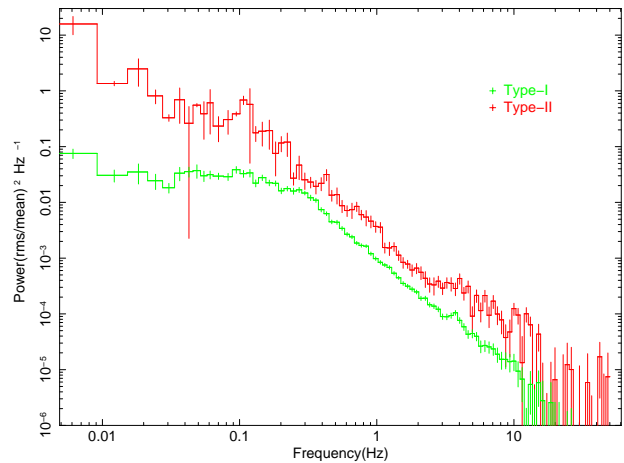}
\label{pds}
\caption{Power density spectrum for {\it type I} and {\it type II} observed on 23rd April 1999. Total power of {\it type II} state is higher than that of {\it type I}.}
\end{figure}

To check the variability of different bands in $\epsilon$ class, we plot RMS amplitude variation in Figure 5
. \citet{Ka02} has showed that rms amplitude of variability increases with energy for a variability system with significant disk emission but the relation is insignificant when disk emission is sufficiently low. Our RMS amplitude diagram clearly show that average RMS variability for $\epsilon$ class is significantly high for high-energy band (14.0-60.0 keV) compared to low energy-band (2.0-8.0 keV) 
in case of {\it type-I}. From spectral analysis we know that {\it type-I} spectra is mostly dominated by disk blackbody component. Hence for $\epsilon$ class, the proposition that presence of black body component in spectra reduces the rms variability in lower energy band is clearly seen.

\begin{figure}
\includegraphics[angle=-90,scale=.350]{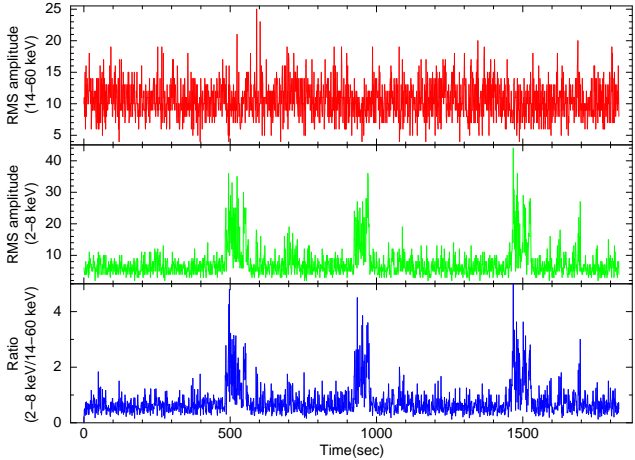}
\label{rms}
\caption{{\it Top Panel:} Variation of RMS amplitude in $\epsilon$ class for 16-60 keV band as observed on 23rd April 1999. {\it middle panel:} Variation of RMS amplitude for 2-8 keV band. In the bottom panel we show ratio of the RMS amplitude for the above two mentioned band. }
\end{figure}
\subsection{Spectral features of the new class}

\begin{table*}
\centering
\caption{Observation ID and Spectrum fitting parameter for {\it Type II}}
\begin{tabular}{ccccccccc}
\hline
Obs. ID&Date&$\Gamma_1$&$T_b$&$\chi^2/f$&$F_{2-26}$&$F_{pow}$&$F_{blackbody}$&$F_{iron}$\\
&&&(keV)&&&&&\\
\hline
40403-01-07-00& 04/23/1999 &3.58$\pm$0.03 &2.06$\pm$0.02 &0.98 &1.15 &0.72 &0.43 &0.002\\
40703-01-13-01& 05/01/1999 &3.65$\pm$0.04 &1.97$\pm$0.02 &0.96& 1.94& 1.26& 0.67& 0.008\\
40703-01-14-00& 05/08/1999 &3.63$\pm$0.04 &1.99$\pm$0.02 &1.00& 1.09 &0.64& 0.44 &0.005\\
40703-01-27-01& 08/23/1999 &3.64$\pm$0.04 &2.02$\pm$0.02 &1.06 &1.06 &0.60 &0.47 &0.004\\
40703-01-27-00& 08/23/1999 &3.71$\pm$0.05 &2.00$\pm$0.02 &1.11 &1.10 &0.62 &0.47 &0.008\\
40703-01-28-02& 08/28/1999 &3.54$\pm$0.04 &2.07$\pm$0.18 &0.98 &1.18& 0.61 &0.57& 0.003\\
40703-01-29-00& 09/05/1999 &3.52$\pm$0.03 &1.98$\pm$0.02 &0.94 &1.32& 0.69 &0.62& 0.004\\
\hline
\end{tabular}
\flushleft Column density $\sim7 \times 10^{22}$ cm$^{-2}$ was used. Flux is given in the unit of photon (1 photon$ = 9.012 \times 10^{-9}$ ergs/cm$^{-2}$/sec in 2-26 keV band).
\end{table*}

The spectral fitting parameters for {\it type I} and {\it type II} is shown in Table 1 and 2. The $90\%$ confidence level for each fitted parameter and $\chi^2/f$ are also shown. The flux of the individual components are also included. In Figure 6
, we have shown different components for {\it type I} and {\it type II} of the spectrum.

\begin{figure*}
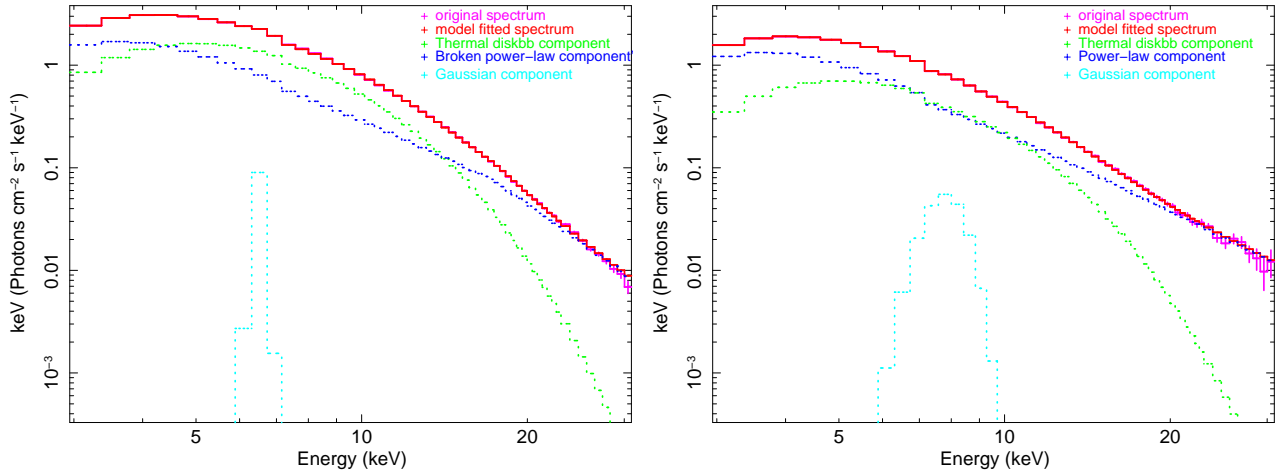

\includegraphics[angle=-90,scale=.350]{figs/high-final.cps}
\includegraphics[angle=-90,scale=.350]{figs/low-final.cps}
\caption{PCA spectrum of GRS 1915+105 in $\epsilon$ class observed on 23rd April 1999. In left panel the spectrum and different model component in {\it type I} region and in right panel spectrum and different model component in {\it type II} region is shown.}
\label{spectrum}
\end{figure*}

From the best fitted values, it is found that {\it type I} spectra can be fitted well with broken power law with a break near 18 keV, along with disk black-body model and small Gaussian at $\sim$ 6.4 keV while all the {\it type II} spectra are found to be fitted with simple power-law, along with disk black body and small Gaussian. The contribution for individual modeling components are also shown in the same figure. For {\it type I} spectra, the contribution of thermal component flux to the total flux is varying between 56\% to 70\% between different observations, while the power-law flux contributed only 26\% to 43\%. Hence, it is clear that it is quite similar to HS (High State) as described by \cite{Ho05}. The break in the power-law continuum suggests that there is two-temperature accretion flow in this mode. In case of {\it type II} spectra, it is found that the hard power law component is contributing 58\% to 66\% of the total flux, while the disk black body is contributing from 40 to 43\% of the total flux. Hence {\it type I} is similar to the LS (Low State) as described by \cite{Ho05}. It is a very important fact to notice the temperature of {\it diskbb} is very high and normalization value is low, suggesting very hot disk and very low inner radius. Hence, accretion rate is very high. We have also tried to fit the spectrum with {\it Comptt} model, but the contribution of compton cloud flux to the total flux is less than 1\% and the optical depth is very low, suggesting the absence of optically thin cloud. Looking at the spectrum point of view, we can conclude that this $\epsilon$ class is basically intermediate transient type of class where the power-law and thermal component have varing relative contributions.

\section{Discussion}
The class is observed when GRS 1915+105 was undergoing rising phase of a X-ray flare though the source was quiet in radio. 
Studying the new class sequentially, it can be concluded that as the X-ray flare intensity is getting higher, the $\epsilon$ class is spending longer time in {\it type I} and lesser time in {\it type II} and gradually the {\it type II} vanishes. The phenomenon inevitably suggested that the $\epsilon$ class is an oscillatory transitive stage between low and high state in the color-intensity evolution diagram and finally get stabilized at hard state. 
This show that at time of $\epsilon$ class, the source was undergoing some thermal viscous instability when it was trying to adjust between hard and soft states when the strong emission from radio jet was not ON and the source is in fully radiative mode.

\section*{Acknowledgments}
This research has made use of data obtained through the High Energy Astrophysics Science Archive Research Center on line Service, provided by the NASA/Goddard Space Flight Center. We used facilities of National Centre of radio Astrophysics.

\bsp

\label{lastpage}

\end{document}